# Vibrational damping effects on electronic energy relaxation in molecular aggregates

Mantas Jakučionis[1], Vladimir Chorošajev[1], Darius Abramavičius[1]

[1]*Institute of Chemical Physics, Vilnius University,*
*Sauletekio Ave. 9-III, LT-10222 Vilnius, Lithuania*

Representation of molecular vibrational degrees of freedom by independent harmonic oscillators, when describing electronic spectra or electronic excitation energy transport, raises unfavourable effects as vibrational energy relaxation becomes inaccessible. A standard theoretical description is extended in this paper by including both electronic-phonon and vibrational-phonon couplings. Using this approach we have simulated a model pigment-protein system and have shown that intermode coupling leads to the quenching of pigment vibrational modes, and to the redistribution of fluctuation spectral density with respect to the electronic excitations. Moreover, new energy relaxation pathways, opened by the vibrational-phonon interaction, allow to reach the electronic excited state equilibrium quicker in the naturally occurring water soluble chlorophyll binding protein (WSCP) aggregate, demonstrating the significance that the damping of molecular vibrations has for the intrarmolecular energy relaxation process rate.

## I. INTRODUCTION

Molecular aggregates are the natural functional materials employed by Nature in complex energy harvesting processes, ubiquitous in primary processes of photosynthesis [1, 2]. Due to resonant interactions between pigment molecules, it becomes possible to transport molecular excitations, originating from light absorption, to energy conversion centers (*e.g.* PSI and PSII reaction centers). In this sense, such molecular aggregates show properties of semiconductor solar cells [3, 4]. However, the dynamic characteristics of the molecular and crystalline semiconductors are totally different: the crystalline materials demonstrate coherent electron and hole propagation, while the molecular aggregates rely mostly on incoherent excitation hopping [5–8].

The poor electronic coherence, or delocalization, is related to the considerable dissipative effects imposed by the phonon bath and intramolecular vibrations [9]. Decay of coherent processes is of a secondary importance for the light harvesting function in solar applications. The primary factor is the quantum yield of energy transport [10], irrespective of the operational mechanism (coherent vs. incoherent). Moreover, molecular vibrations perform a productive function in assuring efficient energy transfer by matching energy levels of different molecules in a fundamentally heterogeneous medium [11]. Simultaneously, quicker vibrational relaxation may increase the overall incoherent energy delivery pace, thus allowing the following energy conversion or chemical energy production steps to occur at a faster rate.

To understand the underlying molecular excitation energy relaxation mechanisms it becomes necessary to develop proper theoretical models of the molecular systems. The standard theoretical approach is that of an open quantum systems [9, 12–14]. In such description the electronic degrees of freedom constitute the observable part of the system, while all nuclear degrees of freedom are treated on the same level as the thermal reservour (bath), *i.e.* the intramolecular and phonon degrees of freedom together compose the bath, represented by an infinite set of *independent* harmonic oscillators. There is a wide range of approximate theoretical approaches (as well as formally exact), which allow to evaluate excitation dynamics in the open quantum systems [15–21].

While these approaches account for the electronic energy transport, the treatment of intramolecular degrees of freedom becomes questionable. Indeed, no energy relaxation or exchange in the bath oscillators is allowed. This especially becomes apparent in the electronic ground state, where molecular vibrations become isolated and their resulting dynamics becomes reversible, the vibrational energy relaxation becomes impossible. However, this contradicts experimental and theoretical studies of vibrational energy dissipation phenomenon [22–27].

In this paper we present a formulation of molecular aggregates by including an interaction term in the Hamiltonian that allow to describe intramolecular vibrational relaxation in an arbitrary electronic state. We demonstrate that such process becomes important in excitation energy transport of a simple excitonic system. In Section II we specify the general model formulation and also present the stochastic time-dependent variational theory that is used to calculate the time evolution of the electronic-vibrational wavepackets. In Section III we present a model of a naturally found molecular aggregate, the water soluble chlorophyll binding protein, and use the presented theory to show how the intramolecular vibrational energy relaxation reshapes the effective spectral density – frequency content of the bath spectral density function "observed" from the perspective of electronic sites. Results are discussed in Section IV. Detailed derivation of the equations of motion is presented in Appendix A and description of coupling between intra and inter-molecular vibrations is given in Appendix B.

## II. THEORY

We consider a generic open quantum system model with two kinds of degrees of freedom (DOF): molecule electronic and vibrational DOF. However, we part from the traditional model by further partitioning the vibrational DOF into two manifolds, reflecting the intramolecular vibrational modes (we denote them by index $q$) and the phonon modes ($p$). Idea of separating system vibrational DOF into few domains is not new and have been previously adopted in surrogate Hamiltonian theories [28, 29]. Each manifold is understood as a seperate set of normal modes. For each vibrational mode the quantum harmonic oscillator is assigned. The electronic DOF are linearly coupled to all vibrational DOF ($q$ and $p$). Additionally, vibrational modes of different manifolds are also linearly coupled to each other to allow energy exchange. Modes in the same manifold do not couple among themselves. Assuming that different electronic oscillators correspond to different molecules (sites), we can write the Hamiltonian by

$$\hat{H} = \sum_n \varepsilon_n \hat{A}_n^\dagger \hat{A}_n + \sum_{n,m}^{m \neq n} V_{mn} \hat{A}_m^\dagger \hat{A}_n$$
$$+ \sum_q \omega_q \hat{a}_q^\dagger \hat{a}_q - \sum_n \hat{A}_n^\dagger \hat{A}_n \sum_q \omega_q g_{nq} \left( \hat{a}_q^\dagger + \hat{a}_q \right)$$
$$+ \sum_p \omega_p \hat{b}_p^\dagger \hat{b}_p - \sum_n \hat{A}_n^\dagger \hat{A}_n \sum_p \omega_p g_{np} \left( \hat{b}_p^\dagger + \hat{b}_p \right)$$
$$- \sum_{q,p} K_{qp} \left( \hat{a}_q^\dagger + \hat{a}_q \right) \left( \hat{b}_p^\dagger + \hat{b}_p \right). \tag{1}$$

Here $\varepsilon_n$ denotes the electronic transition energy for site $n$ (augmented by the reorganization energy due to molecular vibrations and the bath) and $V_{mn}$ is the resonant coupling matrix element between sites $m$ and $n$. $\hat{A}_n^\dagger$ and $\hat{A}_n$ are the Paulionic creation and annihilation operators for an electronic excitation at an $n$-th site, while $\hat{a}_q^\dagger$ and $\hat{a}_q$ are the bosonic creation and annihilation operators for the $q$-th molecular vibrational mode. Simillary, $\hat{b}_p^\dagger$ and $\hat{b}_p$ are the creation and annihilation operators of the $p$-th phonon bath mode. Each mode is characterised by the oscillation frequency $\omega$ (we set $\hbar = 1$), and by the dimensionless linear coupling strength $g_{nx}$ between the $n$-th site and $x = q, p$ -th vibrational mode. This coupling causes modulation of the transition energies for all electronic oscillators. The last term in the Hamiltonian describes linear interaction between the molecular and phonon modes, with coupling strength quantified by the matrix $K_{qp}$. The combined energy of the system is shifted by the total zero-point energy of all vibrational modes $-\frac{1}{2}\sum_x \omega_x$, which makes no difference for the excitation dynamics, summation definted as $\sum_x \equiv \sum_p + \sum_q$.

We include the last term specifically to unlock the possibility of energy exchange between molecular and phonon vibrational modes. For reference, lets consider the electronic ground state. In this case, all terms in the Hamiltonian, which involve $\hat{A}_n^\dagger \hat{A}_n$ vanish and the Hamiltonian of the electronic ground state is purely vibrational

$$\hat{H}_0 = \sum_q \omega_q \hat{a}_q^\dagger \hat{a}_q + \sum_p \omega_p \hat{b}_p^\dagger \hat{b}_p$$
$$- \sum_{q,p} K_{qp} \left( \hat{a}_q^\dagger + \hat{a}_q \right) \left( \hat{b}_p^\dagger + \hat{b}_p \right). \tag{2}$$

This form presents the possibility of energy exchange via vibrational modes due to the last term as has been widely described in the community of open quantum systems [30]. This term is assumed to be small in comparison to the molecular vibrational energies and the leading terms are in the resonant coupling $V_{mn}$ matrix.

Microscopic excitation dynamics are obtained by applying the Dirac-Frenkel variational scheme, outlined in Ref. [31], to a Davydov trial wavefunction

$$|\Psi_\mathrm{D}(t)\rangle = \sum_n \alpha_n(t) \hat{A}_n^\dagger |0\rangle_{el} \prod_x |\lambda_x(t)\rangle, \tag{3}$$

with $x = q, p$ and $|\lambda_q(t)\rangle = \exp\left(\lambda_q(t)\hat{a}_q^\dagger - \mathrm{h.c.}\right)|0\rangle_{vib}$ being the vibrational coherent state (the same expression is used for phonons). $|0\rangle_{el}$ is the electronic ground state, $|0\rangle_{vib}$ is the vacuum state of all vibrational DOF. This method optimizes the parametric wavefunction, so that the deviation from the exact solution of the Schrödinger equation is minimised. The complex parameter $\lambda(t)$ characterizes vibrational wavepacket in the phase space: real part being the expectation value of a coordinate, while the imaginary part - the momentum. This approach allows us to approximately treat a large number (thousands) of vibrational modes explicitly, avoiding Born and Markovian approximations used in the perturbative reduced density operator approach. The excitonic subspace of the parametrized solution admits an arbitrary single-excitation state. Further, the electronic ground state can be included as

$$|\Psi_\mathrm{D}(t)\rangle_\mathrm{g} = |0\rangle_{el} \prod_x |\lambda_x(t)\rangle, \tag{4}$$

if is necessary for spectroscopy simulations. In order to obtain a set of differential equations for the free parameters, the time-dependent variational procedure, as described in Ref. [32] is applied (see Appendix A). Equations for the excited state electronic amplitude read as

$$\dot{\alpha}_n = -\mathrm{i}\alpha_n\varepsilon_n - \mathrm{i}\sum_{m \neq n} V_{mn}\alpha_m + \mathrm{i}\alpha_n \sum_x \omega_q g_{nx}(\lambda_x^\star + \lambda_x), \tag{5}$$

and for the intramolecular and phonon bath vibrational mode amplitudes:

$$\dot{\lambda}_q = -\mathrm{i}\omega_q(\lambda_q - \sum_n |\alpha_n|^2 g_{nq}) + \mathrm{i}\sum_p K_{qp}(\lambda_p^\star + \lambda_p), \tag{6}$$

$$\dot{\lambda}_p = -\mathrm{i}\omega_p(\lambda_p - \sum_n |\alpha_n|^2 g_{np}) + \mathrm{i}\sum_q K_{qp}(\lambda_q^\star + \lambda_q). \tag{7}$$

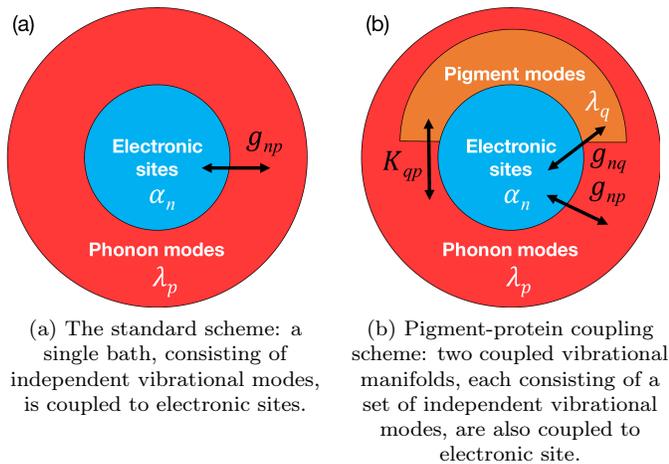

(a) The standard scheme: a single bath, consisting of independent vibrational modes, is coupled to electronic sites.

(b) Pigment-protein coupling scheme: two coupled vibrational manifolds, each consisting of a set of independent vibrational modes, are also coupled to electronic site.

Figure 1: Diagram representation of different coupling schemes between electronic sites and bath used in the description of an open quantum system.

By setting $\alpha_n = 0$ for all $n$, we would get vibrational amplitudes for the electronic ground state.

To capture the main properties of molecular systems we adopt a standard assumption of a local bath – each molecule is surrounded by its own bath. In the current approach it is obtained by assigning sets of phonon modes to specific sites. Electronic oscillators associated with these sites then feel vibrational fluctuations via $g_{nx}$ matrix (here again $x = q, p$). By taking $M$ vibrational DOF per site we need to consider $N \times M$ vibrational DOF for $N$ sites. Accordingly, we define the local spectral density function in terms of electron-vibrational coupling strengths $g_{nx}$ by

$$J_n(\omega) = \frac{\pi}{\omega^2} \sum_x g_{nx}^2 \omega_x^2 \delta(\omega - \omega_x) , \quad (8)$$

which characterize the frequency composition of the vibrational DOF coupled to the $n$-th site. The intramolecular vibrational modes are discrete and countable, while the number of oscillators needed to represent a thermal phonon bath, ideally, is taken to infinity.

The scheme of all interactions is presented in Figure 1. Here we show how the standard description (Figure 1A) is extended by identifying a separate pool of pigment intramolecular modes. Only the $K_{qp}$ coupling matrix binds these manifolds to permit the intramolecular vibrational energy relaxation.

## III. THE INTERPLAY OF INTRAMOLECULAR VIBRATIONS AND PHONONS IN ENERGY RELAXATION

### A. A single optical transition

Effects of linear coupling between vibrational modes will be revealed by numerically simulating a model molecular system. Water soluble chlorophyll binding protein (WSCP) has been suggested as an ideal small system for investigating electronic-vibrational interactions [33]. It contains four identical protein units, each binding only one chlorophyll $a$ (Chl $a$) molecule (or Chl $b$), and pigment molecules are arranged in such a way, as to make two pairs of weakly bound dimers, while pigments inside each of the dimers are interacting much stronger [34].

First, a single pigment-protein unit – one Chl $a$ pigment molecule and the corresponding protein molecule in its close vicininy is taken into consideration. By including more units later, we can incrementally raise the complexity of the model. Hence for primary analysis we study a single electronic transition coupled to the vibrational degrees of freedom. One part of these are the intramolecular vibrations, while the other part - the vibrations of surrounding medium or phonons. A scheme of all different couplings present in a model pigment-protein system is shown in Fig. (1B).

Since each pigment-protein unit in WSCP is identical, and protein molecule vibrations affect only the closest pigment molecule, we can consequently assume that the experimentally measured spectral density function is the same for each of its four units $J_n(\omega) \to J(\omega)$. The low frequency content of $J(\omega)$ is generally attributed to the phonon vibrations, thus we denote it by $J_{\mathrm{ph}}(\omega)$. These vibrations are overdamped, and so the phonon spectral density function lineshape is broad, smooth, and in the case of WSCP, can be approximated [35] by the lognormal distribution function:

$$J_{\mathrm{ph}}(\omega) = \frac{1}{\omega\sqrt{2\pi}} \sum_k \frac{S_k}{\sigma_k} \exp\left\{ -\ln\left(\frac{\omega}{\omega_k^c}\right)^2 / 2\sigma_k^2 \right\} , \quad (9)$$

with a standard deviation $\sigma_k$, cutoff frequency $\omega_k^c$ and Huang-Rhys factor $S_k$ values for $k$-th log-normal component. These parameters are given in Table (I). Reorganization energy of $J_{\mathrm{ph}}(\omega)$ is $\Lambda^{\mathrm{ph}} = 65$ cm$^{-1}$. Simulations were done with the phonon spectral density function represented by 1000 modes spanning the interval $\omega = 0\ldots 1000$ cm$^{-1}$. This setup is accurate enough to yield the converged picture. The coupling amplitudes for the phonon modes follow from the Eqs. 8 and 9:

$$g_p^2 = \frac{\Delta}{\pi} J_{\mathrm{ph}}(\omega_p) , \quad (10)$$

where $\Delta$ is the discretization frequency step. The sign of $g_p$ is taken as positive.

Table I: A) Parameters for approximating the experimentally measured WSCP phonon spectral density function $J_{\mathrm{ph}}(\omega)$. B) WSCP resonant coupling matrix $V_{mn}$ values.

A)

| $k$ | $\sigma_k$ | $\omega_k^c$ | $S_k$ |
|---|---|---|---|
| 1 | 0.4 | 28 | 0.45 |
| 2 | 0.2 | 54 | 0.15 |
| 3 | 0.2 | 90 | 0.21 |
| 4 | 0.4 | 145 | 0.15 |

B)

| $V_{mn}$, cm$^{-1}$ | | | |
|---|---|---|---|
| 1 site | | | |
| 100 | 2 site | | |
| 7 | 15 | 3 site | |
| 15 | 7 | 100 | 4 site |

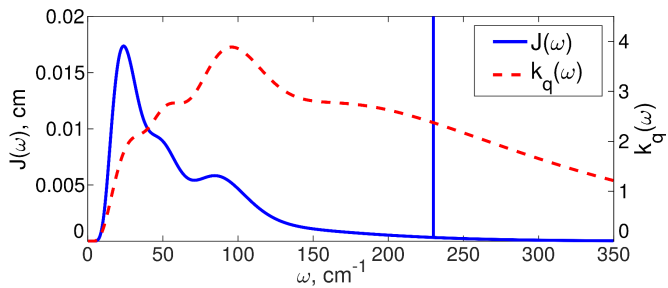

Figure 2: Model spectral density function $J(\omega)$ (blue line) adapted from WSCP pigment-protein complex. It consists of the low frequency phonon modes and a high frequency pigment vibrational mode at $\omega_p = 230$ cm$^{-1}$. The sharp peak is shown schematically – its linewidth is vanishing. V-P coupling strength function $k_q(\omega)$ (dashed red line) calculated with $\gamma = 1$ and $\Delta = 1$ cm$^{-1}$ (see the main text for parameter definitions).

Besides protein vibrations, pigment molecules also vibrate and contribute to the overall modulation of the electronic Hamiltonian. Studies of Chl $a$ using resonance Raman, spectral hole-burning and fluorescence line-narrowing techniques have revealed that a large number of vibrational modes in the frequency range of 150 cm$^{-1}$ < $\omega$ < 1650 cm$^{-1}$ are coupled to electronic sites [36–41]. For now, we take into account only one intramolecular vibrational mode $q = 1$ characterized by the frequency $\omega_q = 230$ cm$^{-1}$, and consider two different electronic-vibrational coupling strength regimes: first is characterised by $g_q = 0.15$, so the corresponding mode reorganization energy $\Lambda^{\text{int}}_{\text{small}} = 5.2$ cm$^{-1}$ is rather small; second being $g_q = 0.33$ with large reorganization energy $\Lambda^{\text{int}}_{\text{large}} = 25$ cm$^{-1}$. Frequency of the chosen mode can with certainty be distinguished as a sharp peak from available difference fluorescence line-narrowing experiment done on WSCP [40]. Reorganization energy values of both regimes are also within the range reported for Chl $a$. Additionally, included mode frequency is within the reach of electronic coupling elements and might be of importance for excitonic energy transport. The resulting full input spectral density function $J(\omega)$ is shown in Fig. (2). It should be noted that experimentally deduced spectral density function is the result of an already coupled molecule and phonon vibration evolution, and, in principle, should not be considered as a separable entity. We will return to this question in section IV.

Next we have to define the $K_{qp}$ matrix which characterizes molecular vibrational damping. For this purpose we assume that the molecular electronic and vibrational DOF interact with the same phonon bath DOF (same phonon modes), so that the $g$ and $K$ matrices can be related. We first define the dimensionless vibrational-phonon (further on we refer to it by V-P) coupling density function via

$$k_q(\omega) = \gamma\omega\sqrt{\frac{1}{\pi\Delta}J_{\text{ph}}(\omega)}, \quad (11)$$

here $\gamma$ is a scaling factor, so that we simply get

$$K_{qp} = k_q(\omega_p)\Delta. \quad (12)$$

Derivation of this form is given in an Appendix B.

For our model system the $k_q(\omega)$ function is displayed by the red dashed line in Fig. (2) when $\gamma = 1$, $\Delta = 1$ cm$^{-1}$. It defines a rather smooth curve at phonon frequencies, meaning that all phonon modes couple with the pigment vibrational mode at a similar strength.

By applying previously layed out notation and assumptions to Eq. (5), (6) and (7) we get differential equations of motion for the single pigment and its surrounding

$$\dot{\alpha} = i\alpha\omega_q g_q\left(\lambda_q^\star + \lambda_q\right) + i\alpha\sum_p \omega_p g_p\left(\lambda_p^\star + \lambda_p\right), \quad (13)$$

$$\dot{\lambda}_q = -i\omega_q(\lambda_q - g_q) + i\sum_p K_{qp}\left(\lambda_p^\star + \lambda_p\right), \quad (14)$$

$$\dot{\lambda}_p = -i\omega_p(\lambda_p - g_p) + iK_{qp}\left(\lambda_q^\star + \lambda_q\right). \quad (15)$$

Equations of motion (13, 14, 15) can be solved analytically in the rotating-wave approximation (by neglecting complex conjugate terms), however, we choose to solve their full form numerically using an adaptive time-step Runge-Kutta algorithm.

Initial conditions for the propagation are taken to correspond to Franck-Condon ground-to-excited state electronic transition at zero temperature. For this purpose it is necessary to evaluate the global potential minima of the vibrational DOF in the electronic ground state, which becomes dependent on $K_{qp}$ matrix. However, if the interation energy of V-P couplings is small compared to the reorganization energy of electronic-vibrational interaction, the perturbation imposed by the V-P interaction is neglibible. In that case one can use $\alpha(0) = 1$, $\lambda_x(0) = 0$ for all modes. Moreover, even when initial condition are shifted (with respect to a global minimum), V-P interaction will cause Rabi-like oscillations that will cancel out for the slower dynamics of electronic excitation. Keeping this in mind we take $\alpha(0) = 1$, $\lambda_x(0) = 0$ for all modes and inspect the V-P interaction energy in every simulation.

Because $|\alpha|^2$ of a single electronic state is time-independent, for now, we solely focus on the effects of V-P coupling on pigment and phonon vibrational modes. For this purpose we consider the energy of specific vibrational modes with respect to electronic excited state $E_x(t) = \omega_x|\lambda_x(t) - g_x|^2$. In this representation $E_x(t)$ should decay to zero in the long time when V-P interaction is infinitesimal. The energy of the pigment vibrational mode ($\omega_q = 230$ cm$^{-1}$), calculated with V-P coupling factor $\gamma = 0 - 0.225$ in the $\Lambda^{\text{int}}_{\text{small}}$ and $\Lambda^{\text{int}}_{\text{large}}$ pigment mode regimes, is displayed in Fig. (3). As expected, when V-P coupling is absent, the energy remains constant. When the V-P coupling is on ($\gamma > 0$), we observe oscillation and dissipation of the vibrational energy. In the regime of $\Lambda^{\text{int}}_{\text{small}}$, energy of the vibrational mode quickly rises just after the initial excitation, afterwards slowly decays to the lower energy value. As the total reorganization energy of the phonon bath is much greater than the vibrational reorganization energy $\Lambda^{\text{ph}} \gg \Lambda_{\text{small}}^{\text{int}}$, the highly oscillatory behaviour is due to pigment vibrational mode intermixing with phonon modes and behaving as a mediator between different phonon modes thus facilitating energy flow among them. Such a large increase in energy is due to the fact, that a single pigment vibrational mode is considered. By including tens of intramolecular modes and keeping V-P coupling energy the same, initial energy increase is distributed among all intramolecular modes and the energy increase per single mode is much smaller (not shown). In the case of $\Lambda_{\text{large}}^{\text{int}}$ the intermixing of the vibrational modes with the phonon modes is not significant and intramolecular vibrational energy dissipation becomes clearly evident: the energy relaxes to a low value at the long time.

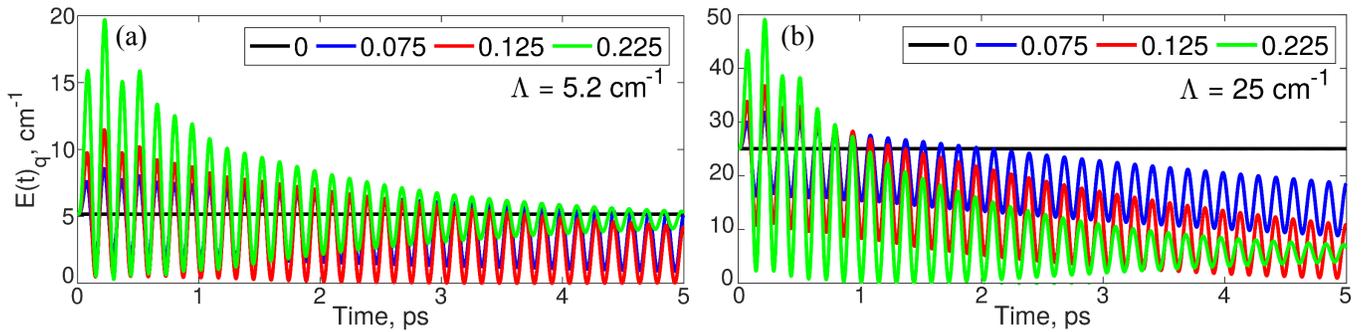

Figure 3: Time dependence of the pigment vibrational mode $E_q(t)$ ($\omega_q = 230$ cm$^{-1}$), obtained with V-P coupling factor $\gamma = 0 - 0.225$ in the small (a, $\Lambda = \Lambda_{\text{small}}^{\text{int}}$), and large (b, $\Lambda = \Lambda_{\text{large}}^{\text{int}}$) pigment mode reorganization energy regime.

It should be noted that when V-P coupling is strong $\gamma = 0.225$, the total energy of V-P interaction becomes comparable, in the case of $\Lambda_{\text{large}}^{\text{int}}$, or larger, in the case of $\Lambda_{\text{small}}^{\text{int}}$, than the electronic reorganization energy. This can be observed by the fact that the green curve in Fig. 3a) decays to a higher value than with weaker V-P interactions. This is not the case in Fig. 3b). Thus, the case of $\gamma = 0.225$ implies the V-P coupling becomes too large, however, we keep this parameter regime for reference.

Energy conservation dictates that changes of pigment vibrational mode energy must alter energy landscape of the rest of the modes. As we are mainly interested in the vibrational energy changes induced by the V-P coupling, and not the absolute values, we calculate time-dependent difference for each of the phonon mode $p$

$$\Delta E_p(t) = E_p(t)_{\gamma \neq 0} - E_p(t)_{\gamma = 0} , \quad (16)$$

so that the positive value indicates an energy increase. A map of $\Delta E_p(t)$, computed with V-P coupling factor $\gamma = 0.075$ and $\gamma = 0.225$ in the regimes of $\Lambda_{\text{small}}^{\text{int}}$ and $\Lambda_{\text{large}}^{\text{int}}$ is presented in Fig. (4). For each regime, the maximum intensity of coresponding plots are cut at the value shown on each of the plots in percents, so that the intensity scale is the same and direct comparison between plots with different $\gamma$ values is possible. Notice that qualitatively both pictures are similar, so the strength of mode intermixing does not play a decisive role.

By first looking at plots with weak $\gamma = 0.075$ and strong $\gamma = 0.225$ V-P couplings in $\Lambda_{\text{small}}^{\text{int}}$ regime, we see that the majority of noticeable energy changes occur for phonon modes of frequency similar and below that of the intramolecular vibrational mode $\omega_q = 230$ cm$^{-1}$. Highly oscillatory, but persistent (in time) picture of energy change for modes $\omega_p < 200$ cm$^{-1}$ suggest that V-P coupling induced energy redistribution for these modes occur almost instantaneously. This is further confirmed by the total phonon vibrational mode energy change $\Delta E^{\text{tot}}(t) = \sum_p \Delta E_p(t)$ plots in Fig. (5) with different $\gamma$ values, obtained by performing integration of Fig. 4 plots along $\omega_p$. We clearly see that higher the V-P coupling strength, more energy is absorbed from intramolecular vibrational mode. However, it takes longer to reach equilibrium for energy difference $\Delta E_p(t)$ of frequency $\omega_p \approx 230$ cm$^{-1}$ phonon modes. These modes show weak oscillations, but gradually gain energy over time, while modes of $\omega_p \approx 250$ cm$^{-1}$ are actually losing energy. This long lasting resonant energy exchange also manifests itself as oscillations visible in $\Delta E_p^{\text{tot}}(t)$ at long times. At 5 ps, more than half of total absorbed energy resides in phonon modes of $\omega_p \approx 230$ cm$^{-1}$. In the $\Lambda_{\text{large}}^{\text{int}}$ regime the dynamics are similar.

V-P coupling induced vibrational mode energy redistribution signify a strong intermixing of various modes. This implies that the spectral density function $J(\omega)$ initially used to characterize electron-vibrational coupling strengths is no longer a proper measure of an actual electron-vibrational coupling strength felt by the electronic DOF of the molecule. The proper spectral density could be obtained by shifting vibrational coordinate system into a new set of non-interacting normal modes. Alternatively, the spectral density could be obtained from the presented dynamics of vibrational degrees of freedom as follows. To deduce the actual properties of coupled system-bath fluctuations, that are felt by the electronic sites, we introduce an effective spectral density function

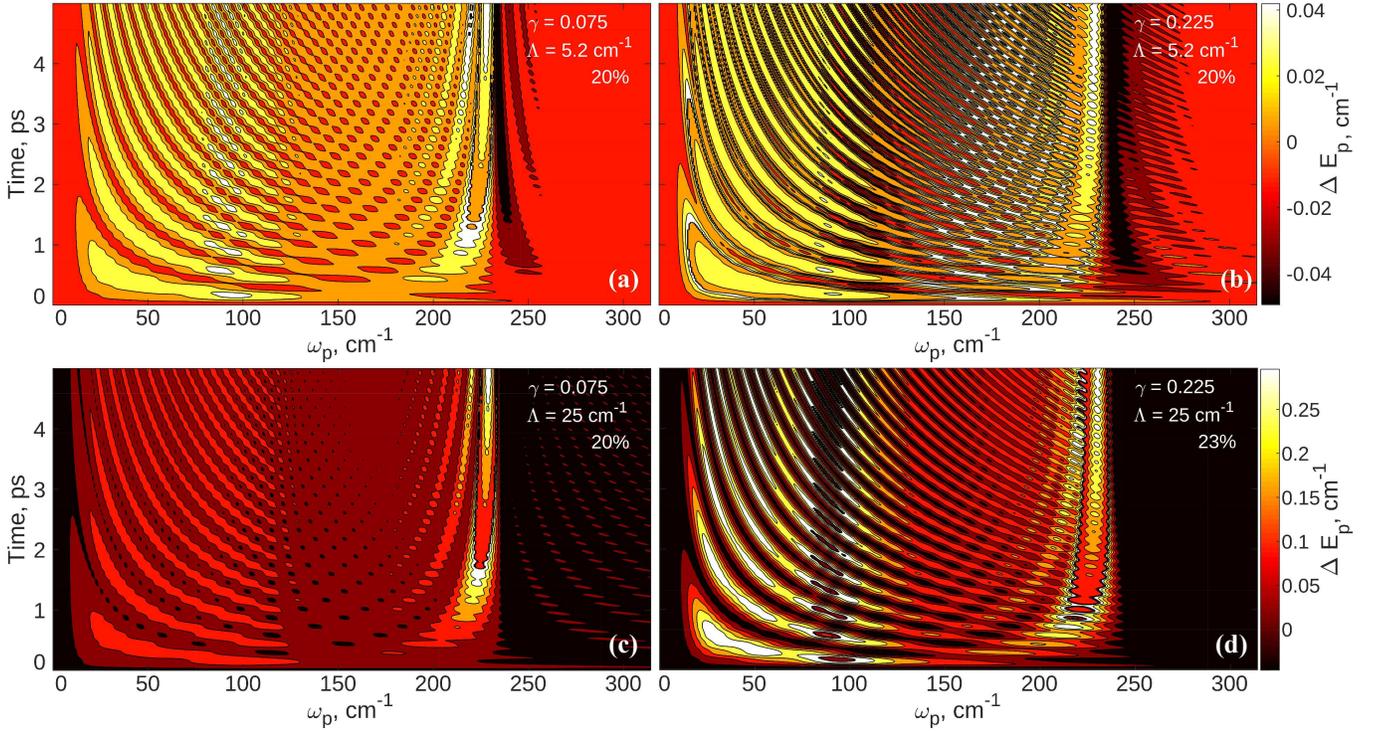

Figure 4: Map of the phonon vibrational mode energy $\Delta E_p(t)$ change in time due to the V-P coupling, calculated with V-P coupling factor $\gamma = 0.075$ and $\gamma = 0.225$ in the small (a), (b), and large (c), (d) pigment mode reorganization energy regime. The maximum intensity of each plot is cut at the value show on in percents.

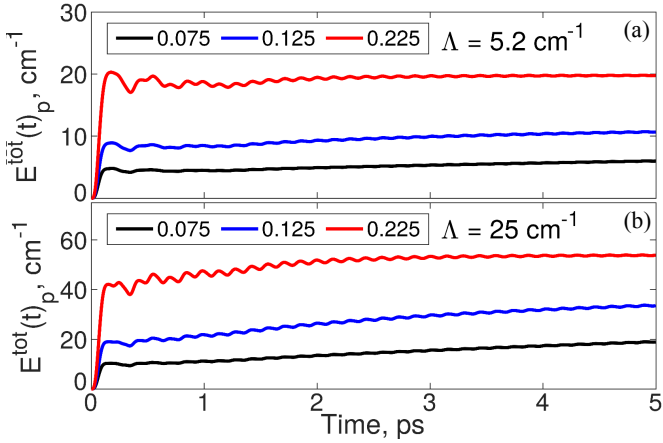

Figure 5: Combined phonon vibrational mode energy change in time $\Delta E_p^{\text{tot}}(t)$, calculated with V-P coupling factor $\gamma = 0.075 - 0.225$ in the small (a), and large (b) pigment mode reorganization energy regime.

$J_{\text{eff}}(\omega)$:

$$J_{\text{eff}}(\omega) = C \cdot \sum_x \left| \tilde{\lambda}_x(\omega) \right|^2, \qquad (17)$$

where $\tilde{\lambda}_x(\omega) = \mathcal{F}\{\lambda_x(t) - g_x\}$ is a Fourier transformed vibrational amplitude oscillation function about its origin point in coordinate-momentum phase space. Index $x$ enumerate both molecular and phonon vibrational modes, and $C$ is the normalization constant chosen so that the total reorganization energy remains unchanged

$$C = \frac{\Lambda^{\text{ph}} + \Lambda^{\text{int}}_{\gamma=0}}{\Delta \sum_\omega \sum_x \omega \left| \tilde{\lambda}_x(\omega) \right|^2 \Big|_{\gamma=0}}. \qquad (18)$$

In Fig. (6) we present an effective spectral density function $J_{\text{eff}}(\omega)$ calculated with V-P coupling factor $\gamma = 0 - 0.225$ in both small and large pigment mode reorganization energy regimes. As expected, when vibrational modes do not interact, $J_{\text{eff}}(\omega)_{\gamma=0}$ perfectly matches the originally used spectral density function $J(\omega)$, previously seen in Fig. (2), this confirms validity of the concept. Independent of the pigment mode reorganization energy regime, V-P coupling induces few effects on a system vibrational degrees of freedom. First, lineshape of a pigment vibrational mode looses most of its intensity, broadens, becomes asymmetric, profile of the broadened peak is neither Gaussian nor Lorenzian (see insets); effects indicative of a vibrational mode quenching phenomenon. Second, lineshape of the low frequency phonon modes intensity increases without drastic changes to its original shape, indicating increased phonon mode oscillation amplitudes; this should have a significant effect on the electronic excitation time evolution as shown below.

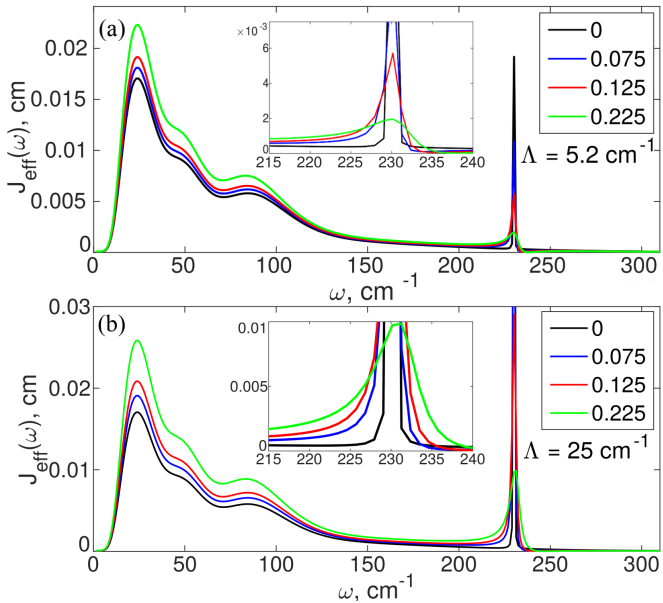

Figure 6: Effective spectral density function $J_{\text{eff}}(\omega)$ calculated with V-P coupling factor $\gamma = 0 - 0.225$ in the small (a) and large (b) pigment mode reorganization energy regime. Insets display $J_{\text{eff}}(\omega)$ close-up of an intramolecular mode.

### B. Excitation dynamics in WSCP aggregate

To study V-P coupling effects on excitation energy transfer rates, we expand the model system and include all 4 pigment-protein units of the WSCP. Since each of 4 electronic sites are identical, we set all excited state energies to $\varepsilon_n = 0$ cm$^{-1}$. Electrostatic interaction strength between different electronic sites are described by the resonant electronic coupling matrix $V$, whose values are given in Table (I). We continue with an assumption that each WSCP pigment is effected only by neighbouring protein vibrations, thus each electronic site is electron-vibrationaly coupled to a local vibrational bath. Each local bath, as in the model system, consists of 1000 phonon vibrational modes of frequency $\omega = 0 \ldots 1000$ cm$^{-1}$, but now we also include 20 pigment vibrational modes of frequencies spanning the interval $\omega = 186 \ldots 742$ cm$^{-1}$ with the combined reorganization energy $\Lambda^{\text{int}} = 93$ cm$^{-1}$ per local bath. So in simulations we have a total of 4080 vibrational degrees of freedom.

We have simulated time evolution of the WSCP aggregate reduced excitonic density matrix $\rho(t)$ in the exciton basis at a temperature of 77 K, whose matrix elements are given by

$$\rho_{e_i e_j}(t) = \left\langle \sum_{n,m} \psi^*_{n e_i} \psi_{m e_j} \alpha^*_n(t) \alpha_m(t) \right\rangle, \qquad (19)$$

$\psi_{m e_j}$ are the eigenbasis vectors of the excitonic Hamiltonian in Table (I) and $\langle \ldots \rangle$ is an ensemble average over a large number of independent trajectories. Molecular aggregate is usually understood as a microscopically disordered system with static energy fluctuations [1, 2, 42] and represent majority of spectroscopically investigated samples. Ensemble averaging was performed over vibrational mode initial conditions $\lambda_x(0)$, determined by the Boltzmann distribution sampling procedure presented in [43] for including a finite temperature; over uniformly distributed direction of excitation electric field $\vec{E}$, which in turn determine initial conditions for electronic sites $\alpha_n(0) = \vec{\mu}_n \cdot \vec{E}$, with $\vec{\mu}_n$ being the ground-excited state transition dipole moment of a pigment molecule $n$, given in [34]; and over electrostatic site excitation energy (diagonal Table (I) elements), sampled from the normal distribution $\mathcal{N}(m, \sigma)$ with mean $m = 0$ cm$^{-1}$ and standard deviation $\sigma = 60$ cm$^{-1}$ values [44, 45].

Time evolution of diagonal $\rho(t)$ elements, which represent the expectation value of finding excitation in the excitonic state of WSCP, calculated without pigment modes; with pigment modes and no V-P coupling ($\gamma = 0$); with pigment modes and V-P coupling ($\gamma = 0.15 - 0.225$) is presented in Figure (7). Excitonic states are arranged in an increasing energy order. Interaction with an external electric field $\vec{E}$ causes system to transition into a non-equilibrium excitonic state, thus $\rho(t)$ dynamic represent equilibration process by which excitation energy is coherently transferred between pigment molecules and is gradually delocalised over several electronic sites, eventually reaching equilibrium between excitonic states. In the case of highly symmetric WSCP, energy becomes delocalised over all four pigments equally, also a third excitonic state initially is the most populated [33].

When no intramolecular modes are included, for 200 fs after interaction with an electric field, energy is being slowly transfered to lower and quickly to higher excitonic states. This rather strange initial upward energy transfer is due to WSCP being made of two weakly interacting dimers. Its excitonic states are arranged into two pairs ($\rho_{11}$, $\rho_{22}$ and $\rho_{33}$, $\rho_{44}$) of states with high energy gap between the pairs and low energy gaps between the states of each pair. This small energy gap causes populated $\rho_{33}$ state to coherently transfer energy upwards during the lifetime of coherence. Nonetheless, after 200 fs the excess energy that the highest excitonic state has gainned from $\rho_{33}$ begins to relax – an increase in population of $\rho_{33}$ is visible at 300 fs. From there on, it takes about 1 ps for equilibrium configuration to be reached, with majority of excitation energy residing in the lowest two excitonic states. When non-interacting ($\gamma = 0$) intramolecular vibrational modes are included, population transfer dynamic is rather unchanged, although energy equilibration is quicker. In the case of coupled pigment vibrational and phonon modes in each of the local baths ($\gamma = 0.15 - 0.225$), energy redistribution occurs even faster, it also change equilibrium state energy distribution. More energy stay in high energy excitonic states, signifying the redistribution of electronic-phonon interactions by inter-mode mixing and deviation from the Boltzmann distribution of the final excitonic states population.

Model timescales of the initially quick and afterwards slow energy relaxation processes qualitatively match those reported in [46]. They measured that WSCP system exhibit ultrafast (50 fs) intra-dimer energy relaxation and slow (1 ps) inter-dimer energy transfer. Quantitative agreement between model and experiment relaxation rates could be achieved by actually optimizing resonant coupling constants between pigments, as we have chosen rather generic values to simply display the effects of V-P coupling; by including all WSCP pigment vibrational modes reported in [40], which should even further speed up intra-dimer energy relaxation; and by tunning V-P coupling strength parameter $\gamma$.

## IV. DISCUSSION

The problem of describing intramolecular vibrations in molecular aggregates often lead to models where a finite number of vibrational quanta is included into the description. Such approach, denoted as one-, two- or n-particle approximation (nPA), depending on the chosen upper limit of the number of vibrational quanta, yield an accurate approach that has been, *e.g.* efficiently used to describe absorption and emission of J aggregates [47], excitation energy transfer in a chlorophyll dimer of the LHCII [48], exciton dynamics inside Fenna–Matthews–Olson complex [49]. The approach, however, is limited to a small number of explicitly included vibrational DOF. Additionally, formally exact methods for multi-mode Brownian oscillator models have been formulated, *e.g.* hierarchy equation of motion [16, 17, 50–52] and real-time path-integral [53, 54] approach, and successfully applied. All these methods regard bath DOF as indirectly observed (which is justified for the majority of systems found in nature) and average them out to obtain the reduced density matrix. In this paper, however, we use an approximate treatment of the problem, and so we are able to include a large number of system and bath vibrational DOF and treat them on an equal footing.

In time-dependent variational approach (TDVA), each oscillator is parameterized by the continous variable $\lambda(t)$ as a function of time and, in terms of access to multi quanta vibrational states, allow to go beyond TPA approach. On the other hand, the treatment becomes quasi-classical since the evolution of each oscillator is described by a single trajectory, though, aspects of quantumness are retained by the representation of coherent states, which obey uncertainty principle in the coordinate-momentum phase space. Additionally, TDVA is advantageous compared to methods, which include a finite number of quanta, since the scaling of the numerical problem with the vibrational degrees of freedom is not exponential, but linear. The limitation of TDVA is built into the ansatz form. In the present approach we use only a single coherent state to represent the vibrational state of an oscillator. As a result, only single-quanta processes are included in the equations of motion (Eq. 13): only $\omega_x$ is associated with $\lambda_x$, hence n-phonon processes with frequencies $n\omega_x$ are excluded. Such processes play a significant role when the electronic-vibrational interactions are strong, *e.g.* in the Förster energy transfer case. However, when the electronic-phonon interactions are weak, as in WSCP aggregate, the presented approach is sufficient. Extending the ansatz by generalized coherent state approach allows to continously improve the level of accuracy for strong electronic-vibrational interactions [55].

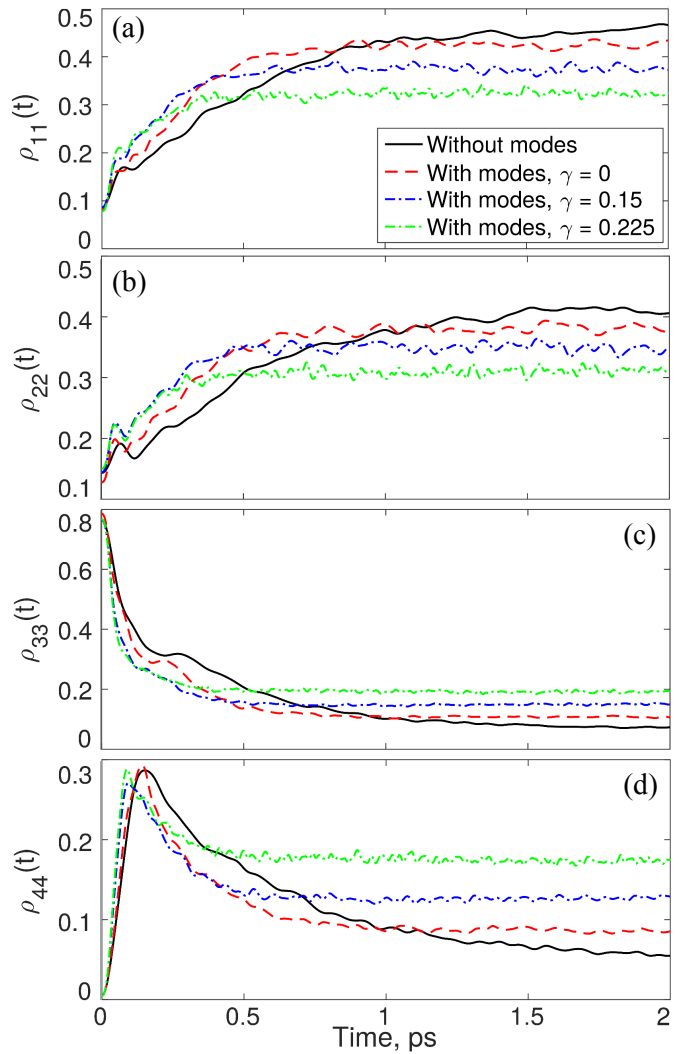

Figure 7: Time evolution of WSCP ensemble excitonic states at temperature of 77 K without intramolecular modes (black line); with intramolecular modes and no V-P coupling ($\gamma = 0$) (dashed red line); with intramolecular modes and V-P coupling ($\gamma = 0.15 - 0.225$) (blue and green dash-dot lines). Excitonic states are arranged in the order of increasing energy.

Majority of theoretical approches fail to capture the ground state intramolecular vibration dissipation, there-fore a proper description of intramolecular vibration energy relaxation is necessary. It appears to be a long-standing and very complicated problem [56]. We assumed a simplified approach by partition all vibrational degrees of system to intramolecular and phonon manifolds, each composed of normal modes, and letting modes of different manifolds interact, as is a common approach in open quantum system theory (*e.g.* Caldera-Leggett model) [30]. By assigning identical phonon spectral densities to electronic energy modulation and to the intramolecular damping, we obtained an analytical expression for the V-P coupling matrix. This is a rather crude approximation, as not necessarily all molecular vibrational DOF must interact with the same bath DOF as does electronic DOF, still, this approach should have provided a rough estimate for the fluctuation distribution accessible to a molecule. For a proper description of initial state we requested that V-P coupling should be small and that all phonon modes are weakly perturbed by molecular vibrations in the electronic ground state, implying that the linear coupling term is sufficient, however, this is still an approximation that forbids thermal heat conductivity in the bulk. During the excitation relaxation the energy is dumped into the phonon modes and it stays there indefinitely. Consequently the heating of specific phonon modes should be taken into account, because it impedes further resonant intramolecular energy dissipation.

There are specific experimental techniques that allow estimation of the phonon spectral density in molecular materials [36–41, 57]. Experimentally derived spectral density accounts for all possible interactions, including V-P couplings. Hence, the measured quantity must show resemblance to our effective spectral density (Fig. 6) as compared to our input function (Fig. 2). The smooth parts of the spectral density stay the same as original input, however, the sharp resonances get smoothed by the V-P couplings. Moreover, this peak broadening must contain temperature dependence due to statistical thermal effects. Consequently, the theoretical input spectral density by definition is temperature independent [9], while temperature dependence in experimentally measured spectral density is a must.

To demonstrate the energy absorption in our model, we additionally present thermal equilibration dynamics of vibrationally hot molecular ground state, with vibrational mode parameters as in the model pigment-protein system (see section III A), by solving Eq. (5), (6) and (7) (the electronic ground state corresponds to electron-vibrational coupling set as $g = 0$ and $\alpha = 0$). We considered intramolecular mode to be effectively at a temperature of 225 K and phonon bath to be at 5 K. Such artificial scenario could be satisfied when electronic energy is dumped into molecular vibrations over a short period of time due to fast internal conversion, and we take such condition merely to demonstrate the very effect. In Fig. (8a) we plot energy differences of the combined phonon bath (solid lines) and intramolecular mode energy (dashed lines) with $\gamma = 0.075 - 0.225$. We clearly see that the energy transfer from intramolecular to phonon vibrational modes is present (part of the energy is accumulated in the V-P coupling interaction energy). In addition, Fig. (8b) shows absorbed energy frequency distribution among phonon modes: most of the energy is resonantly tranfered to $\omega \approx 230$ cm$^{-1}$ phonon modes. Smaller temperature differences between vibrational mode and phonon bath provide, in effect, analogous results.

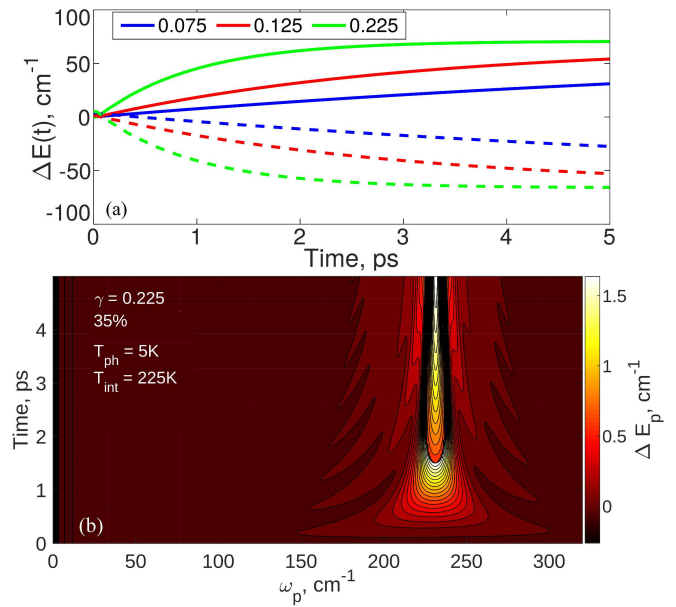

Figure 8: Hot ground state energy differences of the combined phonon bath (solid lines) and intramolecular mode energy (dashed lines), calculated with $\gamma = 0.075 - 0.225$ (a); map of the phonon vibrational mode energy $\Delta E_p(t)$ change in time, calculated with V-P coupling factor $\gamma = 0.225$, and with the maximum intensity cut at 35 percents (b).

Coherent state description submits to an exact treatement of Morse potential (manuscript is being prepared), however, it is computationally expensive, for this reason a more appropriate procedure to tackle the lack of thermal heat conductivity could be to include the nonlinear interaction terms from Taylor expansion of the Morse potential as corrections. In the normal mode basis, due to the Fermi resonances [56], this would lead to heat conductivity in the bath, overall thermal bath equilibration.

We demonstrate that it is rather straightforward to extend the molecular aggregate description using TDVP to include V-P coupling. TDVP with V-P type interaction permit the intramolecular vibrational energy dissipation and redistribution among all vibrational degrees of freedom in both the ground and excited states set in a non-equilibrium configuration.

Finally it should be noted that the V-P coupling has an appreciable effect on the electronic excitation energy relaxation as demonstrated on water-soluble chlorophyll-binding (WSCP) pigment-protein aggregate. We found that the time to reach equilibrium between excitonic states is shorter when V-P coupling is present, it also change equilibrium state energy distribution – population of the two highest energy excitonic states slightly increase, hinting at an excitonic state transformation to vibronic states due to the disturbance by the vibrational couplings.

## ACKNOWLEDGEMENT

This work is financially supported by the Research Council of Lithuania (LMT Grant No. MIP-090/2015).

## Appendix A: Derivation of the equations of motion

To obtain electronic site and vibrational mode amplitude equations of motion using time-dependent variational approach, we begin by calculating Lagrangian $\mathcal{L} = \mathcal{K} - \mathcal{U}$ of the system. Afterwards we use the Euler-Lagrange equations

$$\frac{\mathrm{d}}{\mathrm{d}t}\left(\frac{\partial \mathcal{L}}{\partial \dot{\xi}(t)}\right) - \frac{\partial \mathcal{L}}{\partial \xi(t)} = 0 \, , \tag{A1}$$

where $\xi(t) = \alpha_n^\star(t), \lambda_p^\star(t), \lambda_q^\star(t) \; \forall n, p, q$, to find a set of differential equations $\{\dot{\alpha}_n, \dot{\lambda}_p, \dot{\lambda}_q\}$, which minimize the deviation of the parameterized wavefunction $|\Psi_\mathrm{D}\rangle$ from the exact solution of the Schrödinger equation.

The kinetic part of Lagrangian is equal to

$$\mathcal{K} = \frac{\mathrm{i}}{2}\left(\langle\Psi_\mathrm{D}|\dot{\Psi}_\mathrm{D}\rangle - \langle\dot{\Psi}_\mathrm{D}|\Psi_\mathrm{D}\rangle\right) \tag{A2}$$

$$= \frac{\mathrm{i}}{2}\left(\sum_n \dot{\alpha}_n \alpha_n^\star - \dot{\alpha}_n^\star \alpha_n \right.$$

$$\left. + |\alpha_n|^2 \left[\sum_p \dot{\lambda}_p \lambda_p^\star + \sum_q \dot{\lambda}_q \lambda_q^\star - \mathrm{h.c.}\right]\right) \tag{A3}$$

while calculation of the potential energy term leads to an expression

$$\mathcal{U} = \langle\Psi_\mathrm{D}|\hat{H}|\Psi_\mathrm{D}\rangle \tag{A4}$$

$$= \sum_n |\alpha_n|^2 \varepsilon_n + \sum_{n,m}^{m \neq n} V_{mn} \alpha_m \alpha_n^\star$$

$$+ \sum_n |\alpha_n|^2 \left(\sum_p \omega_p |\lambda_p|^2 + \sum_q \omega_q |\lambda_q|^2\right)$$

$$- \sum_n |\alpha_n|^2 \sum_p \omega_p g_{np} \left(\lambda_p + \lambda_p^\star\right)$$

$$- \sum_n |\alpha_n|^2 \sum_q \omega_q g_{nq} \left(\lambda_q + \lambda_q^\star\right)$$

$$- \sum_{q,p} K_{qp} \left(\lambda_q + \lambda_q^\star\right)\left(\lambda_p + \lambda_p^\star\right) . \tag{A5}$$

By assuming the single-excitation condition $\sum_n |\alpha_n|^2 = 1$, Lagrangian can be simplified to

$$\mathcal{L} = -\mathrm{i}\sum_n \dot{\alpha}_n^\star \alpha_n + \frac{\mathrm{i}}{2}\left(\sum_p \dot{\lambda}_p \lambda_p^\star + \sum_q \dot{\lambda}_q \lambda_q^\star - \mathrm{h.c.}\right) -$$

$$- \sum_n |\alpha_n|^2 \varepsilon_n - \sum_{n,m}^{m \neq n} V_{mn} \alpha_m \alpha_n^\star - \sum_p \omega_p |\lambda_p|^2$$

$$- \sum_q \omega_q |\lambda_q|^2 + \sum_n |\alpha_n|^2 \sum_p \omega_p g_{np} \left(\lambda_p + \lambda_p^\star\right)$$

$$+ \sum_n |\alpha_n|^2 \sum_q \omega_q g_{nq} \left(\lambda_q + \lambda_q^\star\right)$$

$$+ \sum_{q,p} K_{qp} \left(\lambda_q + \lambda_q^\star\right)\left(\lambda_p + \lambda_p^\star\right) . \tag{A6}$$

Inserting this expression into the Euler-Lagrange equations, we obtain the presented Eq. (5) for electronic site and Eq. (6), (7) for vibrational mode amplitudes.

## Appendix B: Pigment-protein vibrational coupling

Stochastic fluctuation-dissipation theorem implies that the coupling between phonon and pigment vibrational motions should lead to the pigment vibrational motion amplitude quenching [9]. Dynamics of $q$-th pigment vibrational motion can be described by the Lengevin equation

$$\ddot{Q}_q(t) + \omega_q^2 Q_q(t) + \int_0^t d\tau \zeta_q(t-\tau) \dot{Q}_q(t) = \xi(t) \, , \tag{B1}$$

with a time-dependent damping kernel [30]

$$\zeta_q(t) = \theta(t) \sum_p \frac{K_{qp}^4}{\omega_p^2} \cos(\omega_p t) \, , \tag{B2}$$

and stochastic force $\xi(t)$ induced by the vibrational motion of phonons. They can be related by the fluctuation-dissipation theorem. First we determine the Fourier transformed time-dependent damping kernel

$$\zeta_q(\omega) = \int \frac{w \, dw}{2\pi} F_q(w) \frac{\eta + \mathrm{i}(\omega - w)}{\eta^2 + (\omega - w)^2} \, , \tag{B3}$$

this additionally defines a dimensionless phonon spectral density function

$$F_q(\omega) = \pi \sum_p \frac{K_{qp}^4}{\omega^2 \omega_p} \delta(\omega - \omega_p) . \tag{B4}$$

Further, by assuming that the pigment electronic and vibrational degrees of freedom interact with the same protein vibrational degrees of freedom, we can define a general phonon spectral density function $J_q(\omega)$ in terms of $F_q(\omega)$ by

$$J_q(\omega_p) = \frac{\omega_p}{\gamma^2 K_{qp}^2} F_q(\omega_p) . \tag{B5}$$

$J_q(\omega)$ is related to V-P coupling matrix $K$. Combining Eq. (B4) and (B5) we obtain

$$J_q(\omega) = \frac{\pi}{\gamma^2} \sum_p \frac{K_{qp}^2}{\omega \omega_p} \delta(\omega - \omega_p), \quad \text{(B6)}$$

by setting $J_q(\omega) \equiv J_{\text{ph}}(\omega)$, as in the main text, we finally get

$$K_{qp} = \gamma g_p \omega_p. \quad \text{(B7)}$$

---